\newcommand{\arcsec}{^{\prime\prime}}

\documentstyle{laa}

\begin{document}
\thesaurus{16.13.1; 16.14.1; 03.05.1}
\title{Pulsars Tracing Black Holes in Globular Clusters}
\author{F.~De Paolis\inst{1,2} \and V.G.~Gurzadyan\inst{3} 
\and G.~Ingrosso\inst{1,2}}
\offprints{F.~De Paolis}
\institute{
Dipartimento di Fisica, Universit\`a di Lecce, Via Arnesano, CP 193, 73100
Lecce, Italy
\and
INFN, Sezione di Lecce, Via Arnesano, CP 193, 73100 Lecce, Italy
\and
University of Sussex, 
Brighton BN1 9QH, UK and
Department of Theoretical Physics, Yerevan Physics Institute, 
Yerevan 375036, Armenia (permanent address) 
}

\maketitle

\begin{abstract}
Results of precise measurements of the periods of pulsars
discovered in the central regions of globular clusters are shown to 
be approaching the capabilities of testing the existence of a central
black hole. For example, in the case of M 15 the available 
data on two pulsars PSR 2129+1210A and PSR 2129+1209D seem to exclude the
existence of a black hole, the presence of which was instead supported 
by recent Hubble Space Telescope data on surface brightness profile 
(Yanny {\it et al.}1994). 
The fluctuations of the gravitational field caused by the stars of the 
system are enough to explain the acceleration observed for both pulsars. 
In the case of the Galactic center, additional data are needed for similar 
definite conclusions.   
\end{abstract}

\section{Introduction}
 
Hubble Space Telescope (HST) observations of globular clusters
provide a unique possibility of revealing the nature of their central 
regions.
Do globular clusters contain central black holes 
or are we dealing with purely stellar dynamical phenomena?
The idea of black holes which might be situated at the centers of globular 
clusters has been seriously debated since the first identification of X-ray 
sources within globular clusters (Bahcall \& Ostriker 1975, Silk \& 
Arons 1975, Newell, Da Costa \& Grindlay 1976). 
Ground based studies on surface brightness and velocity dispersion
profiles were not accurate enough for model independent analysis
of the problem (e.g. Dubath, Meylan \& Mayor 1993, Meylan 1994).     
\begin{table*}
\centering
\caption{Main parameters of two central pulsars in M15 (Taylor, 
Manchester \& Lyne 1993) and two pulsars in 47 Tucanae 
(Robinson {\it et al.} 1995).}
\medskip
\begin{tabular}{lllllll}
\hline
\hline
Name & R. A. & Dec. & $P$ & $\dot{P}$ & $d_{min}$ & $d_{max}$ \\
PSR J & (J2000) & (J2000) & (s) & ($10^{-17}$) & (kpc) & (kpc) \\
\hline
2129+1210A & 21:29:58.247 & +12:10:01.30 & 0.1106647087715 ($\pm 3$) &
$-2.107 ~({\pm 3})$ & 8.9 & 11.1 \\
2129+1209D & 21:29:58.274 & +12:09:59.74 & 0.0048028043457 ($\pm 3$) &
$-1.75 ~({\pm 12})$ & 8.9 & 11.1 \\
\hline
0024-72C & 00:23:50.343 & -72:04:31.46 & 0.00575678001161 ($\pm 1$) &
$-0.00498 ~({\pm 1})$ & 4.1 & 4.9 \\
0024-72D & 00:24:13.877 & -72:04:43.82 & 0.00535757328590 ($\pm 1$) &
$-0.00028 ~({\pm 2})$ & 4.1 & 4.9 \\
\hline
\end{tabular}
\end{table*}
Recent HST observations of M 15 (Yanny {\it et al.} 1994), one of the 
most dense clusters, 
support the theoretically predicted star density slope (Bahcall \& Wolf 1976)  
for stars in the presence of 
a central massive black hole and support the possibility that this 
black hole has a mass of $M_{\rm bh}\simeq 10^3 M_{\odot}$.  
However, the observed surface brightness distribution can also be
explained via a King model with core radius $r_c=2.2 \arcsec$
(Phinney 1993, Gebhardt \& Fischer 1995).   

The aim of the present paper is to suggest the possibility of using 
millisecond pulsars as a trace of the central regions of globular clusters 
with the particular aim of revealing the existence of central black holes.
Millisecond radio-pulsars were discovered
and intensively studied in recent years (see e.g. Taylor, Manchester 
\& Lyne 1993).

Among the eight pulsars discovered in the globular cluster M 15,  two, 
i.e. PSR 2129+1210A and PSR 2129+1209D (also known as PSR 2127+11A and 
PSR 2127+11D), are situated rather close (within $1.1 \arcsec$)
to its center, i.e. inside $0.05$ pc from the center,  
for the distance to M 15 of $10$ kpc (Phinney 1993).   
The main characteristics of these pulsars are given in Table 1, together 
with two more pulsars in 47 Tucanae with negative $\dot P$.

Obviously, we are considering the case of M 15 
Tucanae, not only because of
the recent interpretation supporting the presence of a central black hole 
(Yanny {\it et al.} 1994), 
but also due to the more complete data available at present, both 
on pulsars and the cluster center.

\section{P relation to the (3+1)-Geometry}

The properties of the pulsars' periods $P$ and of their time derivatives 
$\dot{P}$ and the possibility of their accurate measurement
have made pulsars a kind of unique cosmic clock. 
Along with this the present accuracy of the localization of pulsars 
within the globular cluster as well as the resolved centers of the latter, are
approaching the values when the afore mentioned properties of the
pulsars can become tracers of a central black hole and/or of the dynamical
parameters of the cluster.  

Indeed, the relation of the proper time interval $\Delta \tau$ for 
a source located in the central symmetric gravitational field 
with the radial coordinate, yields:
\begin{equation}
\Delta \tau=\sqrt{g_{00}} \Delta t ~. \label{1}
\end{equation}
where $g_{00}=(1-r_g/r)$, $r_g$ being the gravitational radius.
From here, one can easily derive that a pulsar moving in the vicinity 
of a Schwarzschild black hole from the distance $r_1$ to $r_2$ within the time 
interval $\Delta T$ will undergo the following apparent variation of period
\begin{equation}
\frac{\Delta P}{P} \simeq \frac{1}{2} \frac{r_g}{r} \frac{\Delta r}{r}~,
\label{2}
\end{equation}
with $\Delta r \equiv r_1-r_2 = v~\Delta T$, $v$ the pulsar's velocity,
$r$ the mean distance of the pulsar from the central black hole and 
$\Delta r \ll r$. 
Assuming $v\simeq 15$ km s$^{-1}$ 
we obtain the characteristic time scale for the given $\Delta P/P$:
\begin{equation}
\Delta T \simeq 3 ~{\rm yrs} \left(\frac {\Delta P/P}{10^{-12}}\right)
\left(\frac{M_{bh}}{10^3~M_{\odot}}\right)^{-1}
\label{3}
\end{equation} 
for PSR 2129+1210A in M 15. This means that  
the measurements within the time scale (3) performed at any typical epoch 
have to lead to the change of the ratio $\Delta P/P$ at least in the 12th digit.
Here, the value of the  pulsar velocity has 
been chosen to be equal to the stellar velocity dispersion as a 
minimum value following from stellar dynamical considerations; the higher will 
be the proper velocity of the pulsar - the shorter should be the time scale
(3). 
The probable mass range of the central black hole in M15 is obtained
by Yanny {\it et al.} (1994) based on the brightness profile: 
$1\times 10^3 M_{\odot} \leq M_{bh} \leq 3 \times 10^3 M_{\odot}$.
The corresponding gravitational radius of such a black hole is 
$r_g=2GM_{\rm bh}/c^2=2\arcsec \times 10^{-9} (M_{\rm bh}/10^3 M_{\odot})$. 

One can see, that the already observed stability, namely, up to the 12th 
digit, of $P$ for the first pulsar (Table 1) enables us to exclude the 
existence of a 
black hole of mass within the mentioned interval. 

A less restrictive limit is obtained by considering PSR 2129+1209D in M15: 
from the same equation (\ref{3}) we obtain that it is possible to exclude 
the presence of a black hole of a mass greater than $3\times 10^3~M_{\odot}$.

Note that black holes of smaller mass, i.e. of order of $10^2~M_{\odot}$ 
(with numerical factor depending on the parameters of the star cluster), 
are also excluded by considerations of the precise localization in the 
center of the system (Bahcall \& Wolf 1976, Gurzadyan 1982).

What can then be the reason for the negative $\dot{P}$ for both pulsars? 
The fluctuations of the gravitational field
of the stars in the system seem to be sufficient to cause
this effect (Blandford, Romani \& Applegate 1987, Nice \& Thorsett 1992). 
Indeed, for the estimated star density 
$n\sim 10^7$ pc$^{-3}$ (Phinney \& Sigurdsson 1991),
the mean period of fluctuations of the force is $n^{-1/3}/v\approx 3 \times 
10^2$ yrs, where $v\simeq 15$ km s$^{-1}$ is the stellar velocity 
dispersion. If the increase of the velocity dispersion towards the
center of the system measured since
\cite{psc} is confirmed, then this time scale can become even less.
Correspondingly, the lower limit for the acceleration due to perturbations
of the surrounding stars  
should be $\dot{P}/P\simeq 3\times 10^{-17}$ s$^{-1}$, thus exceeding the
observed acceleration parameters for both these pulsars (Table 1);
hence, it should overwhelm the intrinsic spin-down of the pulsar. 

Let us now briefly discuss another two pulsars with negative $\dot 
P$ (see Table 1) in the globular cluster 47 Tucanae, PSR 0024-72C and 
PSR 0024-72D. The distance of these pulsars from the cluster center  
is 1.5 pc and 0.75 pc, respectively, i.e. much higher than the distance of PSR 
2129+1210A and 2129+1209D from the center of M15. 
Therefore, no valuable constraint can be obtained in this case for the 
existence of a central black hole.
Instead, one can use the small value of their $\dot P$ to derive a 
lower limit for the central mass density of 47 Tucanae. 
In fact, due to the mean field acceleration $a_l$ of a pulsar along the 
line of sight, it must be  
(see e.g. Robinson {\it et al.} 1995)
\begin{equation}
\left|\frac{\dot P}{P}\right|<\left|\frac{a_l}{c}\right| ~.
\label{4}
\end{equation}
By using a King model for the mass distribution of the cluster one has,
from equation (\ref{4}),
\begin{equation}
\rho_{\rm min}(0)=\left|\frac{9~c~r}{4\pi G~r_c^2}\frac{\dot P}{P}\right|
~\sim 8\times 10^4~M_{\odot}~{\rm pc}^{-3}~,
\label{5}
\end{equation}
where $r$ is the distance of the pulsar from the cluster's center and 
$r_c\simeq 0.5$ pc is the King {\it core} radius (Calzetti {\it et al.}
1993).

Concerning the situation in the
Galactic center, there are at least two pulsars 
located presumably in its vicinity, i.e. PSR 1748--2446A and 
PSR 1749-3002 (also called PSR 1744--24A and PSR 1746--30), whose main 
characteristics are given in Table 2.  
One of them (PSR 1748--2446A) is located at least 210 pc from the Galactic 
center (taking the distance to the Galactic center to be 10 kpc though the 
real distance can be less (Genzel \& Townes 1987)), so that 
$r_g/r\simeq 5\times 10^{-10}$ for the mass of a central black hole 
$\sim 10^6 M_{\odot}$. 
The available accuracy of measurement of $P$ for that pulsar is again 
of the order of $10^{-10}$, so that it is still not possible to derive 
any definite conclusion on the existence of a black hole in the center of 
the galaxy.

\section{Discussion and predictions}

The accuracy of HST data on the globular cluster centers along
with the measurements of the parameters of radio-pulsars in those
systems can already provide the possibility of studying the
parameters of the stellar systems via the properties of the pulsars.
In particular, these data  can lead to model independent conclusions on the
presence of black holes.
\begin{table*}
\centering
\caption{Main parameters of two pulsars close to the Galactic center 
(Taylor, Manchester \& Lyne 1993).}
\medskip
\begin{tabular}{lllllll}
\hline
\hline
Name & R. A. & Dec. & $P$ & $\dot{P}$ & $d_{min}$ & $d_{max}$ \\
PSR J & (J2000) & (J2000) & (s) & ($10^{-17}$) & (kpc) & (kpc) \\
\hline
1748--2446A & 17:48:02.2534 & $-24:46:37.7$ & 0.01156314838966 ($\pm 2$) &
$-1.9\times 10^{-3} ~({\pm 2})$ & 6.4 & 7.8 \\
1749--3002 & 17:49:13.48 & $-30::02:34$ & 0.60987235659 ($\pm 3$) &
$787.1 ~({\pm 2})$ & 9.11 &  \\
 & & & & & (mean value) & \\
\hline
\end{tabular}
\end{table*}
In the example of M 15 we have seen that the measured parameters of
PSR 2129+1210A (located within $1.1 \arcsec$ from the center), seems to 
exclude the possibility of the existence
of a black hole of mass $\sim 10^3M_{\odot}$, whose existence was instead 
required by the
interpretation of the recent HST photometric data (Yanny {\it et al.} 1994).
Since a black hole of a lower mass is excluded (Bahcall \& Wolf 1976, 
Gurzadyan 1982), the
existence of a black hole in the center of this cluster seems to be excluded 
at all. 

Note, that in future studies, aside from problems associated
with the measurements of the pulsar parameters, one has to consider 
contribution of various effects which can lead to errors in the 
estimation of distances of pulsars from the center of the 
clusters. Then, it will be necessary to take into account 
the accuracies in the alignment of the ephemerides 
dynamical frame with the optical one, the accurate definition of
the cluster center, e.g. via the luminosity profile or
via the dynamical properties (cf Calzetti {\it et al.} 1993), and so on.
However, the conclusion made above for M 15 is not altered  even   
if the pulsar position with respect to the cluster center is known with an 
error as large as $0.5 \arcsec$; in reality errors are 
expected to be much less (Meylan 1994). 

The conditions within the core of M 15 seem to provide 
also the possibility of explaining the acceleration observed for both pulsars 
as due to the field perturbations caused by the surrounding stars. 
This allows us to make the {\it first prediction} concerning the 
single pulsars, i.e. non-members of binary systems:
{\it the acceleration of pulsars should be 
observed with essentially higher probability in globular clusters
and in the Galactic center, than, say, in the disk of the Galaxy}.   

The effect of perturbations caused by the surrounding stars,
in principle, can be empirically
distinguished from the perturbation caused by the regular field of the
cluster core considered by Wolcszan {\it et al.} (1989), Phinney (1991), 
Nice \& Thorsett (1992). The point is that the role of the core should 
become more important towards the boundary of the cluster and will tend to zero
approaching the center, while the stellar contribution, vice versa,
will increase towards the center. 

This leads to our {\it second prediction}: {\it accelerated pulsars should be
situated mainly in the central regions of the globular clusters, rather
than in their peripheral regions}.
Obviously, some probability remains of the pulsars being perturbed by a chance
encounter with a star. The projection effects should also not be neglected
when considering the corresponding probabilities. 
Moreover the existence of anomalously accelerated pulsars  with 
$\dot{P}$ positive not fitting the typical magnetic field action, also 
should be expected in the central region of globular clusters, due to the 
same effect of stellar perturbations.

Concerning the role of stellar perturbations on pulsars one can predict 
one more effect: the change of the orientation of their rotational axes.
The effect of regular or chaotic variation of the rotational
axes is known for the case of the planets (see, Laskar 1994;
Laskar 1995 and references therein) and can be represented as a problem 
of a system with perturbed Hamiltonian: 
\begin{equation}
H(I,\varphi,\epsilon)=H_0(I)+\epsilon H_1(I,\varphi,\epsilon).
\end{equation}
Since the corresponding problem is integrable
only in the case of a single periodic perturbation term (Colombo, 1966),
one can conclude that stellar perturbations in the form represented
by Chandrasekhar and von Neumann (1943) have to lead to chaotic
variations of the pulsar axes. 

Thus, pulsars cannot avoid this effect, even for the lowest values of their
oblateness. The numerical estimation of the obliquity variation,
time scales, etc., can vary within extremely broad ranges, since it crucially 
(non-linearly) depends upon numerous parameters of the problem; 
the corresponding modelling is a topic of comprehensive studies and is
out of the scope of our paper. We only note that
empirical studies can themselves provide information on this effect as well.   

Our {\it third prediction} therefore
reads: {\it pulsars located in the central regions of globular
clusters, in principle, should undergo chaotic variations of their spin axes
and, as a result, can reveal spontaneous appearances and disappearances 
during their observations.}   

These examples indicate:

\noindent
a) the existence of a special subclass of pulsars, i.e.
those situated in globular clusters with properties different 
from those situated  in the Galactic disk or in the halo; 

\noindent
b) the importance of the search of
pulsars in the central regions of globular clusters and of the Galaxy
in revealing the dynamical structure of those  systems, including the
presence of massive black holes.

\section*{ACKNOWLEDGEMENT}
We are grateful to A. Lyne and M. Tavani for 
explanation of observational details and valuable comments.
One of us (V.G.) is thankful to J. Laskar for valuable discussions. 
This work has been in part supported by INFN, Agenzia Spaziale Italiana
(ASI) and The Royal Society.

\end{document}